\documentclass[amsmath,superscriptaddress,prl,twocolumn] {revtex4}
\usepackage{bm}
\usepackage{enumerate}
\usepackage{graphicx}
\usepackage[dvips]{epsfig}
\usepackage{epsf}
\usepackage{xcolor}

\makeatletter
\newcommand{\Rmnum}[1]{\expandafter\@slowromancap\romannumeral #1@}
\makeatletter

\begin{document}
\title{Heavy fermion $\textit{d-f}$ hybrid and the SmB$_6$ low temperature phase}
\author{Anzhelika V. Buskina}
\affiliation{L.D. Landau Institute for Theoretical Physics, 142432, Chernogolovka, Russia}
\affiliation{Moscow Institute of Physics and Technology, 141700, Dolgoprudny, Russia}
\author{Vladimir A. Zyuzin}
\affiliation{L.D. Landau Institute for Theoretical Physics, 142432, Chernogolovka, Russia}
\begin{abstract}
In this Letter we theoretically study physical properties of a model of heavy fermion $d-f$ hybrid. In the studied model two species of fermions have dispersions with different masses, one being much heavier than the other. Hybridization between the fermions at the crossing point of their dispersions doesn't open a true insulating gap leaving a heavy fermion $d-f$ hybrid at the Fermi level. As a result, our theoretical model qualitatively explains experiments on the low-temperature phase of the SmB$_6$. These are the linear in temperature specific heat, saturation of the resistance, and frequency dependence of the optical conductivity. Calculated optical conductivity shows a broadened peak at the twice the value of hybridization as well as a low-frequency tail.  
\end{abstract}
\maketitle

SmB$_6$ is a material that, at temperatures lower than $40\mathrm{K}$, becomes insulating (for a review, see \cite{RosaFisk}).
The insulating regime occurs due to the hybridization \cite{Anderson1961,KeldyshKopaev1965,Hewson} of conducting $d$-fermions with localized $f$-fermions.
Experiments indeed observe activational behavior of the resistance as a function of temperature \cite{RosaFisk}.
However, the resistance of this material saturates at temperatures lower than about $4\mathrm{K}$ instead of increasing to infinity, as is expected for an insulator.
This fact indicates that there are some charge carriers in the low-temperature phase of SmB$_6$, despite it showing strong features of an insulator.
It has been suggested that SmB$_6$ is a topological insulator \cite{SovietTI1} with gapless edge states, which are responsible for the saturation of resistance at low temperatures \cite{DzeroSunGalitskiColemanPRL2010}.
ARPES experiments \cite{ARPES2013} observe evidence of the existence of edge states in SmB$_6$.
Experimental measurements of the specific heat \cite{Kasuya1979,McQueen2014,Wakeham2016} showed that the specific heat at very low temperatures is linear in temperature, suggesting that it originates from conducting fermions \cite{LandauLifshits}.
However, \cite{Wakeham2016} showed that the dominant contribution to the specific heat comes from the bulk of the sample rather than from the surface.
Experimental measurements of the optical conductivity \cite{Gorshunov1999,Flachbart2001,Hudakova2004,Laurita2016} observe the hybridization gap, but in addition suggest \cite{Laurita2016} that the main contribution to the optical conductivity originates from the bulk of the sample.
Furthermore, experimental measurements of the de Haas–van Alphen (dHvA) effect in the purest SmB$_6$ samples \cite{Tan2015} show the appearance of a giant temperature peak in the temperature dependence of the oscillation amplitude.
This giant peak occurs at temperatures lower than $0.5\mathrm{K}$, and if one fits this peak by the Lifshits–Kosevich formula \cite{LifshitsKosevich} for the oscillation amplitude $A(T)$, where $T$ is temperature,
$
A(T) = 2 \pi^2 \frac{T}{\omega{\mathrm{B}}}/\sinh\left(2 \pi^2 \frac{T}{\omega{\mathrm{B}}} \right),
$
where $\omega_{\mathrm{B}} = \frac{eB}{mc}$ and $B$ is the magnetic field and $m$ is the fermion mass, then one obtains that the mass of fermions responsible for the peak is $m \simeq 100 m_{\mathrm{e}}$ \cite{ZyuzinJETPLett2024,ZyuzinPRB2024,CommentA}. In addition to this, the experiment \cite{Tan2015} observed that the value of the resistance plateau at low temperatures measured at $B=45\mathrm{T}$ decreased compared to that at $B=0\mathrm{T}$. This fact argues against a significant contribution of edge states to the residual resistance of SmB$_6$.

All in all, despite the insulating features that SmB$_6$ shows, there is strong evidence that there are charge carriers in the bulk, called the in-gap states, of SmB$_6$ which are responsible for the saturation of resistance \cite{RosaFisk}, the linear-in-temperature specific heat \cite{Kasuya1979,McQueen2014,Wakeham2016}, the optical conductivity \cite{Laurita2016}, and the giant temperature peak in the dHvA effect \cite{Tan2015}, all taking place at very low temperatures.
Despite significant theoretical effort \cite{DzeroSunGalitskiColemanPRL2010,CurnoeKikoinPRB2000,VarmaPRB,ColemanReview,PalPRB2019} to explain the physics of SmB$_6$, the nature of these charge carriers remains unclear to this day.
In this Letter we propose a simple theoretical model that captures all of the experimental facts discussed above.

%-----------------------------------------------------
%-----------------------------------------------------
%--------------------------------------------------
\textit{Model}.- 
In \cite{ZyuzinJETPLett2024} it has been suggested that a giant temperature peak in the dHvA oscillation amplitude \cite{Tan2015} as well as other peculiarities of \cite{Tan2015} can be theoretically obtained in the model of hyridized $d-$ and $f-$ fermions in which both fermions have dispersion. We note that in all of the previous theoretical models \cite{DzeroSunGalitskiColemanPRL2010,CurnoeKikoinPRB2000,VarmaPRB,ColemanReview} $f-$ fermions were considered to be localized without any dispersion. 

In this Letter we wish to understand whether the theoretical model of \cite{ZyuzinJETPLett2024} can explain other experimental features of SmB$_6$ discussed above. 
The Hamiltonian of the model written in the basis of $\phi_{\mathrm{d}}$ and $\phi_{\mathrm{f}}$ fermions, which form a spinor
$\psi^{\dag} = \left[ \phi^{\dag}_{\mathrm{d}} ,~ \phi^{\dag}_{\mathrm{f}} \right]$, is given by
\begin{align}\label{modelB}
\hat{H} = \int_{\bf k} \psi^{\dag}_{\bf k}\left[\begin{array}{cc}\xi_{\bf k} & \theta \\  \theta^{*} & \alpha\xi_{\bf k} \end{array} \right]\psi_{\bf k} ,
\end{align}
where $\xi_{\bf k} = \frac{{\bf k}^2}{2m} - \mu$ is the dispersion of $d-$ fermions, $\alpha \xi_{\bf k}$ is the dispersion of $f-$ fermions, and $\int_{\bf k}(..) \equiv \int \frac{d{\bf k}}{(2\pi)^3}(..)$. Positive constant $\alpha$ is assumed to be very small, $0<\alpha \ll 1$. In that way $f-$ fermions have a large mass. Parameter $\mu$ is not the Fermi level but rather a value which defines depth of the dispersions. 
Any other possible representation of the two dispersions will be inevitably reduced to the $\xi_{\bf k}$ and $\alpha\xi_{\bf k}$ representation, which conveniently captures the crossing point of the spectra.
Fermi level is chosen to be zero in our model. Parameter $\theta$ is the hybridization between fermions. Although, hybridization in realistic SmB$_6$ is odd in momentum, without the loss of generality, it is chosen in the $s-$wave symmetric form for simplicity of estimating specific heat and conductivity. We will comment on that one more time in the discussion part of the Letter. 

\begin{figure}[t] 
\begin{tabular}{c}
\includegraphics[width=0.5 \columnwidth ]{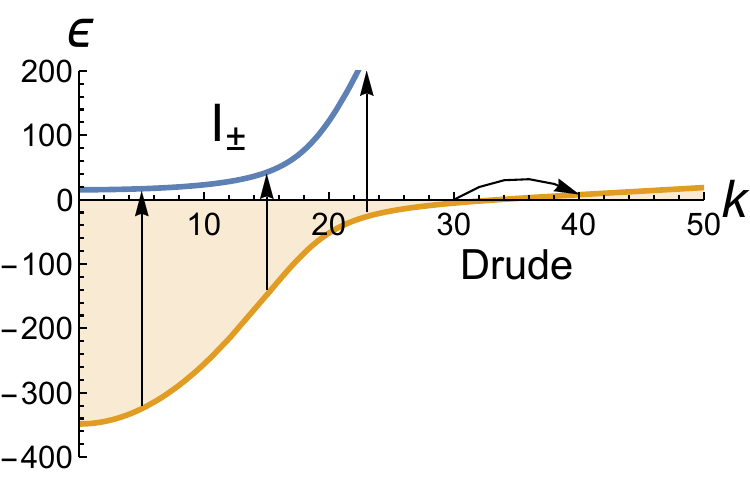} ~~
\includegraphics[width=0.45 \columnwidth ]{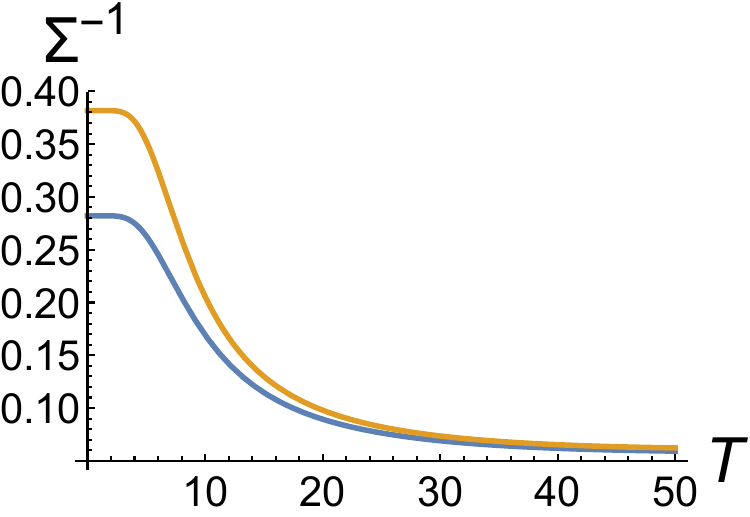} 
\end{tabular}

\protect\caption{Left: dispersion Eq. (\ref{spectrum}) of the model Eq. (\ref{modelB}) for $\theta = 80 \mathrm{K}$, $\mu = 330 \mathrm{K}$ and $\alpha = 1/100$, and schematics of the processes contributing to the conductivity Eq. (\ref{conductivity}) shown by the black arrows. 
Right: resistance as a function of $T$ (in K). Numerical integration of the exact expression of the resistance. 
Parameters are $\theta = 80 \mathrm{K}$, $\mu = 330 \mathrm{K}$, $\tau = 0.003 \mathrm{K}^{-1}$ (recall that $\tau \propto \alpha$), 
Blue for $\alpha=1/100$ and yellow for $\alpha = 1/200$,
such that $\left(\mu+\frac{\theta}{\sqrt{\alpha}} \right) \tau \approx 4.5$ for blue, and $3.45$ for yellow.
}

\label{fig:fig2}  
\end{figure}

The spectrum (shown in Fig. (\ref{fig:fig2})) of fermions after the hybridization is (we assume $\theta = \theta^{*} > 0$) 
\begin{align} \label{spectrum}
\epsilon_{{\bf k};\pm} = \frac{1+\alpha}{2}\xi_{\bf k} \pm \sqrt{\left( \frac{1-\alpha}{2}\xi_{\bf k}  \right)^2 + \theta^2},
\end{align}
with corresponding eigenvectors 
\begin{align}
\psi_{{\bf k}; +} = \frac{1}{N_{\bf k}} \left[ \begin{array}{c} A_{\bf k} \\ \theta\end{array}\right],~~
\psi_{{\bf k}; -} = \frac{1}{N_{\bf k}} \left[ \begin{array}{c} -\theta \\ A_{\bf k}\end{array}\right],
\end{align}
where $A_{\bf k} = \frac{1-\alpha}{2}\xi_{\bf k} + \sqrt{\left( \frac{1-\alpha}{2}\xi_{\bf k}  \right)^2 +  \theta^2}$ is defined in order to facilitate the notations, and $N^2_{{\bf k}} = A_{\bf k}^2 +   \theta^2$ is the normalization.
Rotation matrix is $\hat{T}_{\bf k} = [\psi_{{\bf k};+},~\psi_{{\bf k};-}]$ such that
$
\hat{T}_{\bf k}^{-1}\hat{H}_{\bf k}\hat{T}_{\bf k} =\text{diag}\left[\epsilon_{{\bf k};+}, ~ \epsilon_{{\bf k};-} \right],
$
is a diagonal matrix.
We specifically choose $\theta > \sqrt{\alpha}\mu$ throughout the Letter such that $\epsilon_{{\bf k}; +}$ fermion band isn't occupied when the Fermi level is at zero. On the other hand, $\epsilon_{{\bf k};-}$ band always has a heavy fermion on the Fermi level, whose Fermi surface is defined by $ \xi_{\bf k} = \theta/\sqrt{\alpha}$.

%-----------------------------------------------------
%-----------------------------------------------------
%--------------------------------------------------
\textit{Specific heat.}- Let us first understand temperature behaviour of the specific heat in the model. 
The free energy of the system is 
\begin{align}
F=-T\sum_{\pm}\int_{\bf k}\ln\left[1+e^{-\frac{\epsilon_{{\bf k};\pm}}{T}}\right] + \frac{\theta^2}{U},
\end{align}
where $U$ is the repulsive interaction. Although there is a temperature dependence in $\theta$, for example a mean-field type $\theta(T)=\theta_{0}\sqrt{1-T/T_{\text{c}}}$ for $T<T_{\text{c}}$ behavior, where $\theta_{0}$ is the value of hybridization at $T=0\text{K}$, one can verify that $\frac{\partial \theta}{\partial T}\frac{\partial F}{\partial \theta} = 0 $, that is due to the self-consistency equation, drastically simplifies the calculation. At very low temperatures, when only the $\epsilon_{{\bf k};-}$ fermion band is occupied and $\theta$ is barely temperature dependent, we apply Sommerfeld expansion to get an understanding of how specific heat depends on temepature. We get
\begin{align}\label{HeatCapacity}
c_{\mathrm{v}}^{-} = \gamma T,
\end{align}
where 
\begin{align}\label{gamma}
\gamma = \frac{\pi^2}{3}\left( \frac{1+\alpha}{\alpha}\right)\sqrt{\mu + \frac{ \theta }{\sqrt{\alpha}}},
\end{align}
This is the first main result of this Letter. Namely, due to the presence of the heavy fermion $d-f$ hybrid on the Fermi level, the specific heat at low temperatures, $T < \alpha\left( \mu + \frac{ \theta }{\sqrt{\alpha}} \right) $ behaves linearly with temperature. Coefficient $\gamma$ is enhanced by $\alpha^{-1} \gg 1$ factor. This is consistent with the experiments \cite{Kasuya1979,McQueen2014,Wakeham2016}. At larger temperatures it is expected that phonons would contribute to the specific heat with terms cubic in temperatures. Thus, discontinuity of the heat capacity at the transition temperature might not be experimentally visible.

%-----------------------------------------------------
%-----------------------------------------------------
%-----------------------------------------------------

%-----------------------------------------------------
%-----------------------------------------------------
%-----------------------------------------------------
\textit{Electric conductivity.}- Let us now analyze temperature dependence of electric conductivity of the model.
Longitudinal electric conductivity is given by
\begin{equation}
\sigma(\omega_{m})= e^2\left[ \Pi(\omega_{m})-\Pi(0) \right]/\omega_{m},
\end{equation}
where $e$ is the electron's charge, and
\begin{equation}\label{Pi}
\Pi(\omega_{m})=T\sum_{\epsilon_{n}}\int_{\bf k}\text{Tr}\left[G^{\text{M}}(\epsilon_{n}+\omega_{m};{\bf k})\hat{v}_{\alpha} G^{\text{M}}(\epsilon_{n};{\bf k})\hat{v}_{\alpha} \right],
\end{equation}
is the current-current correlation function. Here $G^{\text{M}}(\epsilon_{n};{\bf k})$ is the Matsubara Green function written in the diagonal basis, and it will be specified below after the disorder averaging is discussed. Matsubara frequencies are, as usual, fermionic $\epsilon_{n} = \pi T (2n+1)$ and bosonic $\omega_{m} = 2\pi T m$. Velocity operator $\hat{v}_{\alpha} =\hat{T}_{\bf k}^{-1} \left( \partial_{\alpha} \hat{H}_{\bf k} \right) \hat{T}_{\bf k}$ has a matrix structure.
%-----------------------------------------------------------------------------------------------------------------
%-----------------------------------------------------------------------------------------------------------------
We add impurities and assume that only metallic fermions from $\epsilon_{{\bf k};-}$ band scatter on them. In the diagonal basis $\hat{H}_{\mathrm{imp}} = \tau_{-} V_{\mathrm{imp}}({\bf r})$, where $2\hat{\tau}^{\pm} = \hat{\tau}_{0} \pm \hat{\tau}_{z }$ are projectors.
Upon treating impurity scattering within the Born approximation we have for the Green function, 
$
G^{\mathrm{M}}(\epsilon_{m};{\bf k}) = \mathrm{diag}[G_{+}^{\mathrm{M}}(\epsilon_{m};{\bf k}) , ~ G_{-}^{\mathrm{M}}(\epsilon_{m};{\bf k}) ],
$
where $G_{+}^{\mathrm{M}}(\epsilon_{m};{\bf k}) = \left[ i\epsilon_{m} - \epsilon_{{\bf k};+} + i0\mathrm{sign}(\epsilon_{m} ) \right]^{-1}$ and $G_{-}^{\mathrm{M}}(\epsilon_{m};{\bf k}) =\left[ i\epsilon_{m} - \epsilon_{{\bf k};-} + \frac{i}{2\tau}\mathrm{sign}(\epsilon_{m}) \right]^{-1}$,
where $\frac{1}{2\tau} = \frac{1+\alpha}{\alpha}\pi v_{\mathrm{imp}}^2n_{\mathrm{imp}}\nu_{3D}$, where $v_{\mathrm{imp}}$ and $n_{\mathrm{imp}}$ is the impurity potential amplitude and concentration correspondingly, and $\nu_{3D} = \frac{m}{2\pi^2}\left[ 2m(\mu+ \frac{\theta}{\sqrt{\alpha}}) \right]^{1/2}$ is the density of states. Note that $\tau \propto \alpha$. The $\text{Tr}$ in Eq. (\ref{Pi}) stands for the trace over the space of $\pm$ fermion bands.

We get for the conductivity
\begin{equation}\label{conductivity}
\sigma(\omega = 0)=\frac{2e^{2}\nu_{3D}}{3\pi m }\left(I_{\mathrm{Drude}}+ I_{\pm} \right) \equiv \frac{2e^{2}\nu_{3D}}{3\pi m } \Sigma,
\end{equation}
where $I_{\mathrm{Drude}}$ is the Drude conducitivity originating from the heavy fermion on the Fermi level, 
\begin{align}\label{Drude}
I_{\mathrm{Drude}} \approx \frac{\pi \alpha\tau}{2} \left( \mu+ \frac{\theta}{\sqrt{\alpha}} \right)^{3/2},
\end{align}
where $\left( \mu+ \frac{\theta}{\sqrt{\alpha}} \right)^{3/2}$ parameter is fixed by the number of fermions in the system. 
It is typical to think that results obtained from the Kubo formalism for the non-interacting fermions treated within the Born approximation don't have any temperature dependence. It is true for $I_{\mathrm{Drude}}$ contribution. In the system under study the hybridization between fermions results in temperature dependence coming from the interband contributions $I_{\pm}$ to the current-current correlation function.
\begin{align}
I_{\pm} &= - 4\pi \tau \theta^2    (1-\alpha)^{2} \int_{-\mu}^{\infty}  d\xi_{\bf k} \frac{\left(\xi_{\bf k}+\mu\right)^{3/2}}{\left[ (1-\alpha )\xi_{\bf k} \right]^{2}+(2\theta)^2}
\nonumber
\\
&\times \frac{1}{1+4 \tau^{2}\left\{ \left[ (1-\alpha) \xi_{\bf k} \right]^{2}+(2\theta)^2 \right\} } \left.\frac{\partial n_{F}}{\partial\epsilon}\right|_{\epsilon=\epsilon_{{\bf k};+}}.
\end{align}
The temperature dependence is due to the Fermi-Dirac distribution function $n_{\mathrm{F}}(\epsilon) =( e^{\epsilon/T} + 1 )^{-1}$ 
in the expression above. At large temperatures the derivative of the distribution function is a smooth function of $\epsilon$, and $I_{\pm}$ contributes significantly to the conductivity. While at low temperatures the derivative becomes exponentially suppressed, and $I_{\pm}$ contribution to the conductivity vanishes. At strictly $T=0$ only the Drude part of the heavy fermion $d-f$ hybrid contibutes, see Eq. (\ref{Drude}). The conductivity of the $d-f$ hybrid is proportional to $\alpha^2$ (recall that $\tau \propto \alpha$ as well as $m_{\mathrm{df}} \propto m/\alpha$) and is thus very small.
We plot numerical estimates of the resistance as a function of temperature in Fig. \ref{fig:fig2}.  
The value of the gap is $2\theta = 160 \mathrm{K}$ (for example, \cite{RosaFisk}), parameter $\mu$ was taken from \cite{Tan2015}, and $\alpha$ was extracted from the position of the large temperature peak \cite{ZyuzinJETPLett2024} of the dHvA oscillations amplitude observed in \cite{Tan2015}.
This is the second main result of the present Letter.

The hybridization parameter $\theta(T)$ is expected to vanish when the temperature is increased. In the limiting case of $\theta = 0$, the conductivity of the system will be given primarily by the light fermion band in Eq. (\ref{modelB}), i.e. $\epsilon_{\bf k} = \xi_{\bf k}$. The conductivity of the system is then given by regular Drude formula with $m$ and $\tau_{0}$ parameters. Thus, the resistance of the system at zero temperature is larger than that at large temperatures by at least $1/\alpha^2 \approx 10^4 $ orders of magnitude. Such a difference isn't conveyed in the Fig. (\ref{fig:fig2}) because plots there are made for $\theta$ being independent on temperature, since we were interested in the low temperature phase only.

What is important is that there is no activational behavior of the resistance of the intuitively anticipated $\propto e^{-\theta/T}$ form. Such an activational behavior occurs only for the case of a true insulator with $\alpha = - 1$. In our case of finite and positive $\alpha$, the activational behvior is much more complicated, and we think there is no easy way to extract the value of the gap from experimental behavior of the resistance with the temperature.

\begin{figure}[t] 
\begin{tabular}{c}
\includegraphics[width=0.45 \columnwidth ]{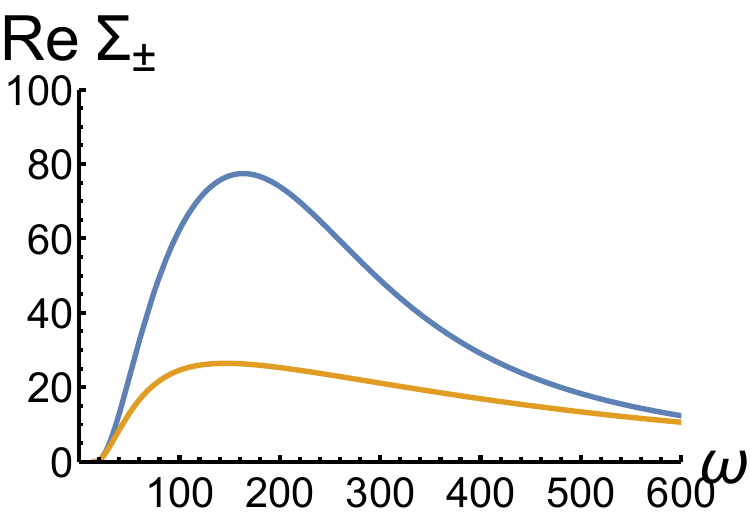} ~~
\includegraphics[width=0.45 \columnwidth ]{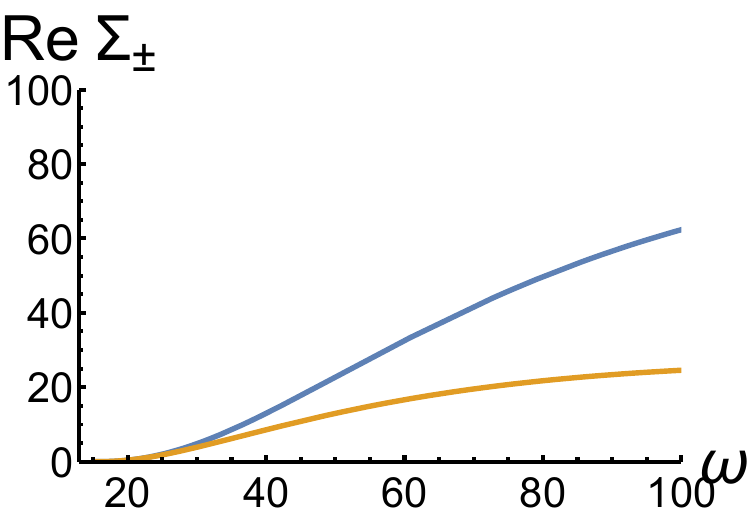} 
\end{tabular}

\protect\caption{Optical conductivity of the interband transitions as a function of frequency (in K) for $T=0$. Parameters are chosen to be $\alpha = 1/100$, $\theta = 80\mathrm{K}$, $\mu = 330 \mathrm{K}$. 
Blue $\tau = 0.003\mathrm{K}^{-1}$ and yellow $\tau = 0.001\mathrm{K}^{-1}$. 
The peak at $\omega = 2\theta = 160\mathrm{K}$ shown in the left plot is smeared by the impurity scattering. A plot in the right emphasizes the low-frequency tail of the optical conductivity. 
It does appear looking like the experimental curves of Ref. \cite{Laurita2016} at low-temperatures.  
Recall that $1\mathrm{THz} \approx 4.8 \mathrm{meV} \approx 48 \mathrm{K}$.
}

\label{fig:fig3}  
\end{figure}

%-----------------------------------------------------
%-----------------------------------------------------
%-----------------------------------------------------
%-----------------------------------------------------
%-----------------------------------------------------
%-----------------------------------------------------
\textit{Optical conductivity.}- 
Let us now study the optical conductivity at $T=0$. Frequency dependence of the interband electric conductivity is
$
\sigma_{\pm}(\omega)=\frac{2e^{2}\nu_{3D}}{3\pi m }\left[ I_{\pm}(\omega)-I_{\pm}(0) \right]/\omega 
\equiv \frac{2e^{2}\nu_{3D}}{3\pi m }\Sigma_{\pm}(\omega),
$ where real part of the conductivity is defined by
\begin{align}
\Re I_{\pm}(\omega)&=4\pi \tau \theta^2 (1-\alpha)^{2}
\int_{-\mu}^{\infty}d\xi_{\bf k}\frac{\left(\xi_{\bf k} +\mu\right)^{3/2}}{\left[ (1-\alpha )\xi_{\bf k} \right]^{2}+(2\theta)^2}
\nonumber
\\
&\times
\frac{\Theta\left(\omega-\epsilon_{{\bf k};+}\right)}{1+ 4\tau^2 \left\{\sqrt{\left[ (1-\alpha)\xi_{\bf k}\right]^{2}+(2\theta)^2}-\omega \right\}^{2} },
\end{align}
where $\omega > 0$ always, and where $\Theta$ is the Heaviside function. Optical conductivity is plotted in Fig. (\ref{fig:fig3}) and it is the third main result of the present Letter. 
As shown in the left plot, there is a peak at $\omega =2\theta$ in optical conductivity. This is reminiscent of the inulating behavior. 
However, unlike in true insulators, the peak is broadened by the impurity scattering life-time $\tau$ of the heavy fermion $d-f$ hybrid.  
In addition, we checked that if $\tau \rightarrow \infty$, the peak would have started almost abruptly from $\omega = 2\theta$.
Furthermore, as emphasized in the right plot, the frequency dependence starts from the minimum of $\epsilon_{{\bf k};+}$ dispersion, i.e. from $\omega = \left. \epsilon_{{\bf k};+} \right|_{\xi_{{\bf k}}= -\mu}$ (approximately $15\mathrm{K}$ for the parameters in Fig. (\ref{fig:fig3}). This can surve as an experimental tool to extract certain details of the spectrum (for example, $\mu$). The broadening shown in Fig. (\ref{fig:fig3}) is the low-frequency tail, and it is consistent with the experimental observations of the optical conductivity \cite{Gorshunov1999, Laurita2016}. For example, in \cite{Laurita2016} optical conductivity of SmB$_6$ at low frequencies $\omega < 2 \mathrm{THz}\approx 96 \mathrm{K}$ was experimentally observed (Fig. 2c and 2d in \cite{Laurita2016}). Their low temperature curves, called anomalous and originating from the in-gap states, are consistent with the right plot of Fig. (\ref{fig:fig3}).

%-----------------------------------------------------------------------------------------------------------------
%-----------------------------------------------------------------------------------------------------------------

%-----------------------------------------------------
%-----------------------------------------------------
%-----------------------------------------------------
\textit{Discussion.}- 
Our analysis above shows that a theoretical model of a heavy fermion $d-f$ hybrid, originally proposed in \cite{ZyuzinJETPLett2024} to understand giant tempearature peak in the dHvA oscillations observed in SmB$_6$ in \cite{Tan2015}, is also consistent with a number of other key experiments on this material. The central in the model is the existence of a heavy fermion $d-f$ hybrid in the Fermi level in the bulk. This hybrid represents the long-sought in-gap state needed to account for experimental observations, whose nature had previously remained unclear.
The model naturally explains three key experimental features: the linear-in-temperature specific heat, the saturation of electrical resistance at low temperatures, and the frequency-dependent optical conductivity.

One might propose that for $ -1 \ll  \alpha <0$ in Eq. (\ref{modelB}), the system behaves as a true insulator, with the Fermi level lying somewhere near the top/bottom of the valence/conduction band (see, e.g., \cite{PalPRB2019,Comment}). If this were the case, however, one would expect that in some sample, either intrinsically or under external influence, the Fermi level could fall within the insulating gap. Such samples would then exhibit resistance diverging to infinity at low temperatures. Yet no such behavior has been observed in SmB$_6$, and all samples display resistance saturation at low temperatures.

As far as the topological edge states \cite{SovietTI1} are concerned, it has been shown in \cite{ZyuzinJETPLett2024} that if the hybridization in Eq. (\ref{modelB}) is odd in momentum, then there are going to be edge states even if the system isn't insulating. 
For example, they are similar to the edge states in Rashba \cite{ZyuzinSilvestrovMishchenkoPRL2007}, and in other systems \cite{DyakonovKhaetskii1981,StanescuGalitski2006}. Such edge states are intertwined with the bulk heavy fermion $d-f$ hybrid, aren't the independent fermion states when thermodynamics and kinetics is concerned, and thus aren't expected to contribute seprately from the bulk states to the free energy and transport. However, it is of interest to theoretically understand how the electric current distributes in the sample \cite{GlushkovJETPLett2022}.
Moreover, it appears that the experiment \cite{Tan2015} suggests that the low temperature phase of SmB$_6$ isn't the topological insulator. Indeed, application of the magnetic field would break the topological protection of the edge states. Then residual $T=0$ resistance, if it was only due to the edge state, will be rising with the magnetic field. However, \cite{Tan2015} observed a slight decrease of the residual resistance with the field.

As far as the two-gap structure of the low temperature phase of SmB$_6$ is concerned, our model can't explain the second, smaller, gap at $2\theta \approx 30-50  \mathrm{K}$. 
For example, the smaller peak is observed in \cite{Hudakova2004} at $2\theta \approx 42  \mathrm{K}$, and in \cite{Laurita2016} (a small hump at about $\omega \approx 0.75 \mathrm{THz} \approx 36\mathrm{K} $ visible in Fig. (2c) in \cite{Laurita2016}) in the optical conductivity measurements.
We have checked that for $\mu \approx 330 \mathrm{K}$ such $\theta$ can't result in a pronounced width of the plateu of the residual resistance occurring below $4\mathrm{K}$. Only the hybridization parameter consistent with the larger experimentally observed gap (of $2\theta \approx 160-200 \mathrm{K}$) describes well the width of the plateau shown in Fig. (\ref{fig:fig2}). In addition, the low-frequency tail in the optical conductivity observed in \cite{Laurita2016} (Fig. 2c) is also consistent with the larger gap (shown in Fig. (\ref{fig:fig2})). Perhaps, the smaller gap corresponds to some intricate degeneracy splitting \cite{Fisk2016,Ropka2019} of the heavy fermion $d-f$ hyrbid bands, and shows up only in the optical conductivity. In this way, the low-frequency tail in  Fig. (2c) in \cite{Laurita2016} can be a sum of the dependence shown in Fig. (\ref{fig:fig3}) and a similar one due to the smaller gap. This is beyond the model studied in this Letter.

%-----------------------------------------------------
%-----------------------------------------------------
%-----------------------------------------------------
\textit{Conclusions.}- 
All in all, a theoretical model of the heavy fermion $d-f$ hybrid proposed and analyzed in this Letter qualitatively describes a number of key experimental, and so far unexlpained, features of the SmB$_6$. Thus we can conclude that our arguments strongly suggest that SmB$_6$ isn't a true insulator but rather a system with a heavy fermion $d-f$ hybrid on the Fermi level in the bulk. The hybrid has both metallic and insulating features. Metallic part is responsible for linear in temperature specific heat Eq. (\ref{HeatCapacity}), saturation of resistance Fig. (\ref{fig:fig2}) and low-frequency tail of the optical conductivity, right Fig. (\ref{fig:fig3}), all three occurring at low temperatures. Insulating part of the hybrid is responsible for growth of the resistance at intermediate temperatures Fig. (\ref{fig:fig2}) and in the peak in the optical conductivity, left Fig. (\ref{fig:fig3}).

\textit{Acknowledgements.}- The authors are thankful to I.S. Burmistrov, M.M. Glazov, P.D. Grigoriev, and A.S. Mel'nikov for useful discussions.
VAZ is greatful to Pirinem School of Theoretical Physics for hospitality. The work was supported by RSF 25-22-00819.

\end{document}